\documentclass[twocolumn]{aastex631}
\usepackage{bm}
\usepackage{amsmath}
\graphicspath{{./}{figures/}}

\newcommand{\RGM}[1]{{#1}}
\newcommand{\SL}[1]{{#1}}

\begin{document}

\title{A radial limit on polar circumbinary orbits from general relativity}

\author{Stephen Lepp}

\author{Rebecca G. Martin}

\author{Anna C. Childs}
\affiliation{Nevada Center for Astrophysics, University of Nevada, Las Vegas, 4505 S. Maryland Pkwy., Las Vegas, NV 89154, USA}
\affiliation{Department of Physics and Astronomy,University of Nevada, Las Vegas, 4505 S. Maryland Pkwy., Las Vegas, NV 89154, USA}

\begin{abstract}
 A particle orbiting a misaligned eccentric orbit binary undergoes nodal precession either around the binary angular momentum vector (a circulating orbit) or around a stationary inclination (a librating orbit).  In the absence of general relativity, the stationary inclination is inclined by $90^\circ$ to the binary angular momentum vector (aligned with the binary eccentricity vector) and does not depend on the particle semi-major axis.   General relativity causes apsidal precession of the binary orbit. Close to the binary, the behaviour of the particle is not significantly affected, a librating particle precesses with the binary. However,  we find that the stationary inclination and the minimum inclination required for libration both increase with the particle semi-major axis. There is a critical radius beyond which there are no librating orbits, only circulating orbits, and therefore there is a maximum orbital radius for a stationary polar orbiting body. The critical radius is within planet forming regions around binaries with semi-major axis $\lesssim 1\,\rm au$. This has implications for the search for misaligned circumbinary planets and the radial extent of polar circumbinary disks.
 
\end{abstract}


\section{Introduction} \label{sec:intro}


Many planets have been detected orbiting around binary star systems \citep[e.g.][]{Doyle2011,Orosz2012}. All of the currently observed systems are in orbits that are close to coplanar to their host binary orbit. This may be representative of nature \citep{Li2016,MartinD2019} but it may be a result of selection effects since the Kepler binaries are all around very short orbital period binaries.   Transits for even slightly misaligned circumbinary planets are less frequent  than for circumstellar planets because of their nodal precession, however, the probability of a transit at some point is higher \citep{MartinD2015,MartinD2017}. Eclipse timing variations may provide a way to detect highly misaligned planets in the future \citep{Zhang2019}.

We expect that planets may form in misaligned orbits around wider binaries (with semi-major axis $\gtrsim 0.5\,\rm au$) since protoplanetary gas disks around these binaries may be more highly misaligned \citep{Czekala2019}. Observations show that misaligned circumbinary gas disks are common \citep[e.g.][]{Chiang2004,Capelo2012,Brinch2016}. 
HD~98800 was the first circumbinary gas disk to be found in a polar configuration \citep{Kennedy2019}. In a polar alignment, a low mass disk (or planet) is misaligned by close to $90^\circ$ to the central binary orbit with its angular momentum aligned to the eccentricity vector of the binary. This is a stable configuration \citep{Aly2015,Martin2017,Lubow2018,Zanazzi2018,Cuello2019}. 



The dynamics of misaligned circumbinary planet orbits have been studied in depth both analytically  and numerically \citep{Verrier2009,Farago2010, Doolin2011,Zanardi2017,Chen2019}.  Around a circular orbit binary, the nodal precession is always around the binary angular momentum vector, we call these {\it circulating} orbits. However, around an eccentric binary, a sufficiently misaligned orbit can precess instead about the stationary inclination that is aligned with the binary eccentricity vector, we call these {\it librating} orbits. Most previous work has ignored the effects of general relativity that causes an additional apsidal precession of the binary orbit in the same direction as orbital motion \citep{Naoz2017}. \cite{Zanardi2018} examined the effects of GR on the orbit of an asteroid around a star with an inner eccentric orbit planet.  \cite{Childs2021} found that  general relativity (GR) does not affect the dynamics of very close in (orbital radius $r<4\,\rm au$) terrestrial planets around a $0.5\,\rm au$ semi-major axis binary since the planet orbits precess with the binary and remain polar aligned.  

In this work, for the first time, we consider how the additional precession due to GR affects the dynamics of circumbinary planet orbits in wide orbits around close binaries.  Without GR, circumbinary orbital dynamics do not change much with the planet semi-major axis \citep[e.g.][]{Doolin2011}. However, in Section~\ref{calc} we use $n$-body simulations to show that GR significantly alters the orbits of wider orbit planets relative to the binary orbit. We find that there is a critical radius outside of which the orbits are circulating for all initial inclinations. In Section~\ref{critical} we show how this critical radius depends upon the the properties of the binary. In Section~\ref{conc} we discuss the implications of our results for both the dynamics of circumbinary planets and the evolution of circumbinary disks.



\begin{figure}
\includegraphics[width=\columnwidth]{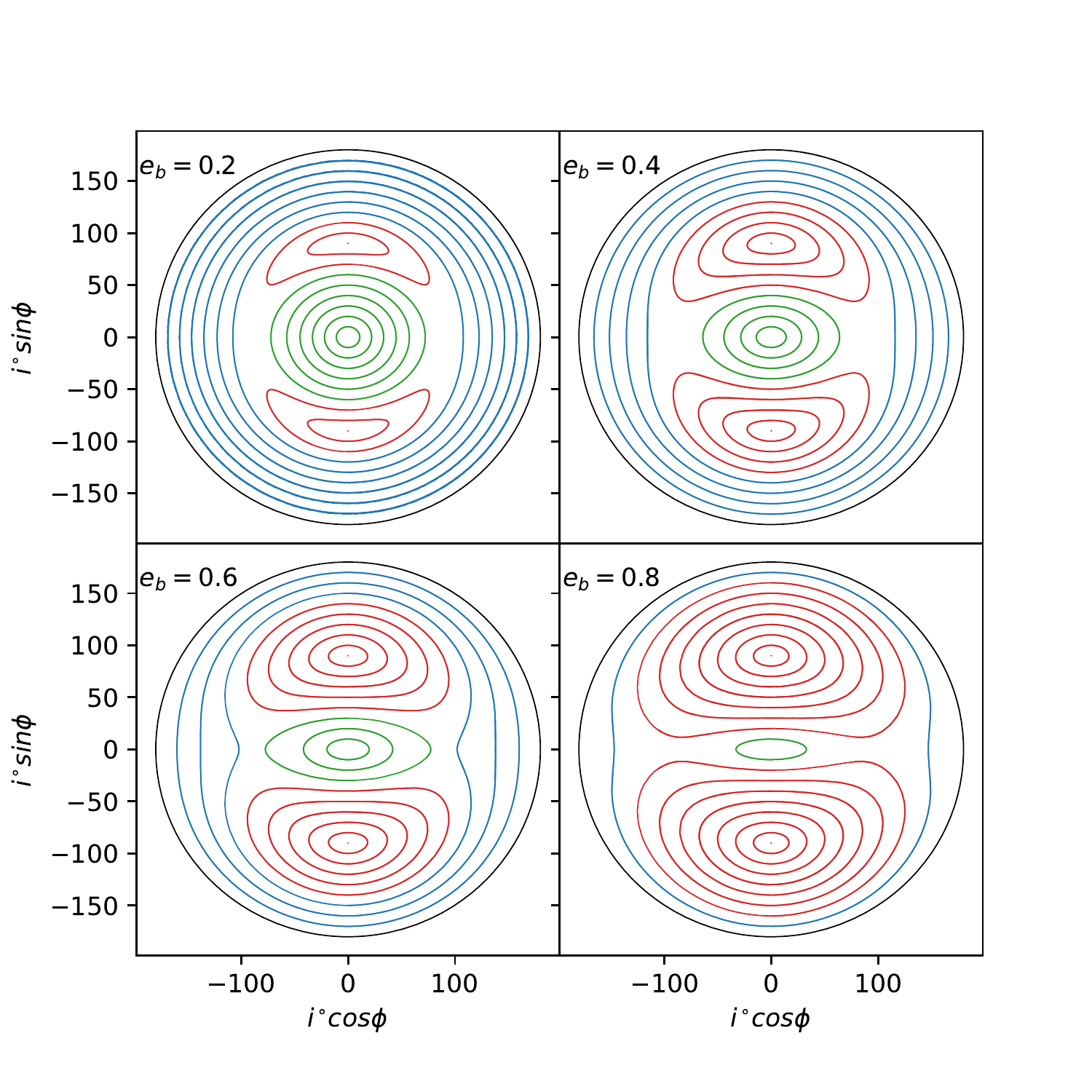}
\caption{The ($i \cos\phi$, $i\sin\phi$) phase diagram 
without GR for a test particle orbiting a binary of two 0.5 
M$_\odot$ stars with a semi-major axis of
$a_{\rm b} = 0.5\,$au and eccentricities $e_{\rm b} = 0.2$ (top left), 0.4 (top right), 0.6 (bottom left) and $0.8$ (bottom right).
The test particles begin in circular orbits at $r=5 \,$au and at initial
inclinations from $i_0=10^\circ$ to $170^\circ$ in increments of
$10^\circ$. Librating orbits are shown in red, circulating 
orbits with inclinations $\leq 90^\circ$ are shown
in green and circulating orbits with initial inclinations $>90^\circ$ are
shown in blue.  The inclination of  $180^\circ$ is shown in black.
}
\label{fig:nogr}
\end{figure}

\section{Circumbinary planet orbital dynamics with GR}
\label{calc}

\begin{figure*}
\includegraphics[width=2.1\columnwidth]{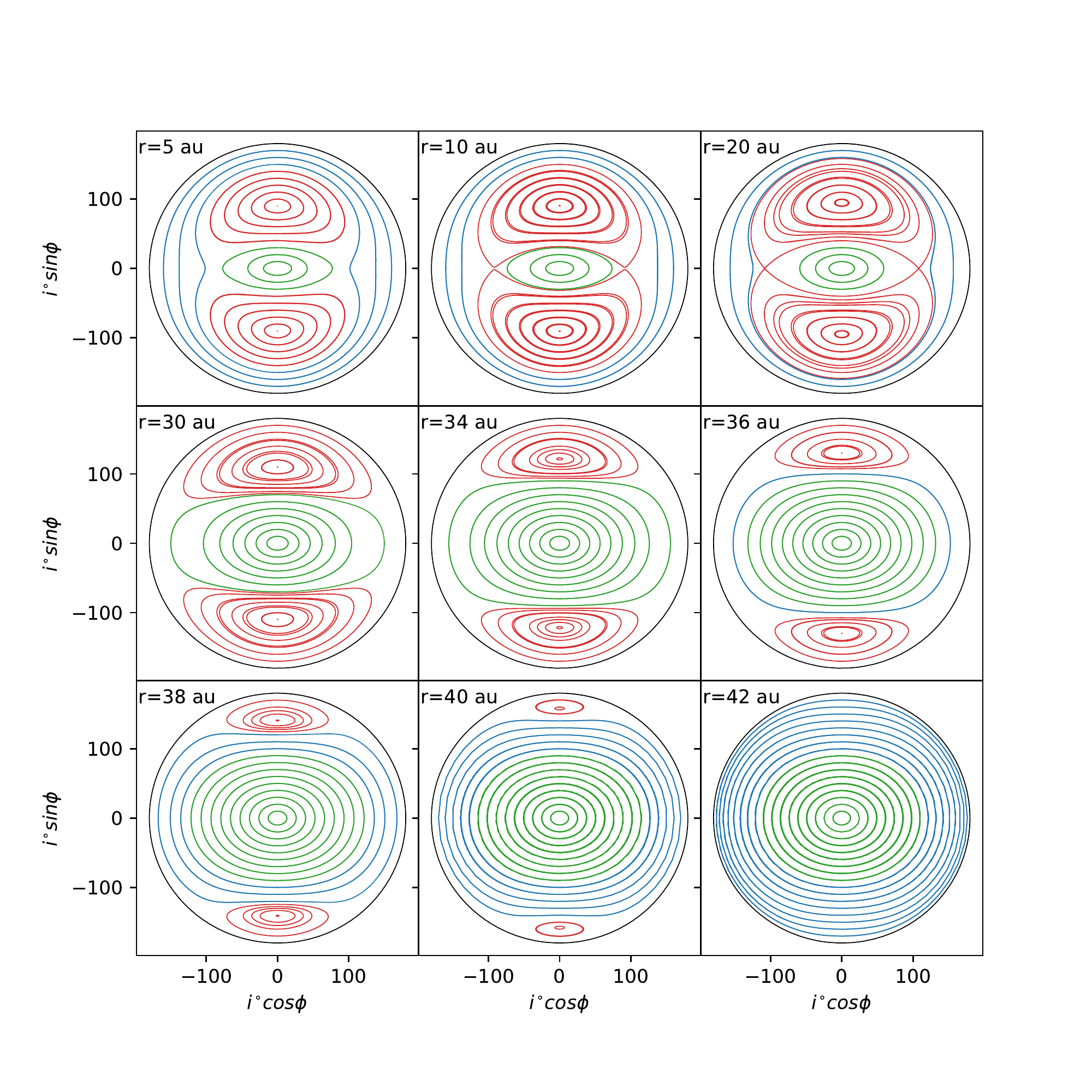}
\caption{The  ($i \cos\phi$, $i\sin\phi$) phase diagram with GR for a binary with 
$a_{\rm b} = 0.5\,\rm au$ and for $e_{\rm b} = 0.6$ for
test particle semi-major axes of $r=5$, 10, 20, 30, 34, 36, 38, 40  and $42\,$au. 
The initial inclinations and color coding as same as fig~1.
}
\label{fig:withgr1}
\end{figure*}

\begin{figure*}
\includegraphics[width=2.\columnwidth]{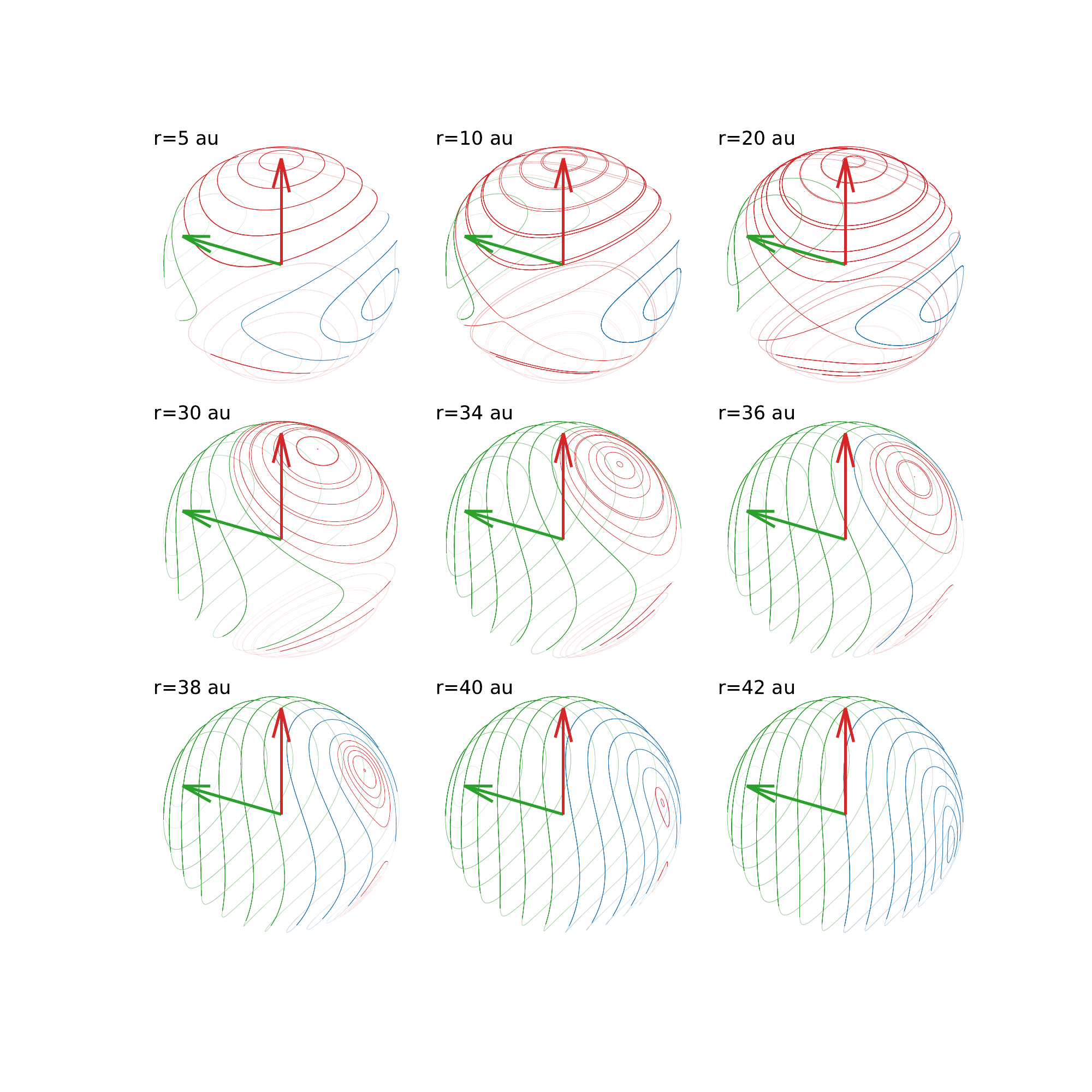}
\vspace{-2cm}
\caption{\SL{The same simulations as Fig~\ref{fig:withgr1} but with the inclination and
phase angle plotted on the surface of a sphere.  The red vector is the binary eccentricity unit vector and the green the binary angular momentum unit vector.
The initial inclinations and color coding are the same as Fig~\ref{fig:withgr1}.}
}
\label{fig:withgr2}
\end{figure*}

The calculations for this paper were made using the
{\sc REBOUND} $n$-body code \citep{rebound} and the {\sc REBOUNDx} package extension to 
include GR effects using the gr\_full package
\citep{reboundx,newhall83}. All simulations used the WHFAST integrator, a symplectic Wisdom-Holman integrator \citep{reboundwhfast,wh},
but the critical points were checked with the  IAS15, a 15th order
Gauss-Radau integrator \citep{reboundias15} and the 
SimulationArchive format was used to have a fully stored
state of the system for analysis \citep{reboundsa}.

We analyse the particle orbits in the frame of the binary. The inclination is the 
angle between the direction of the angular momentum vector 
of the test particles 
orbit and the direction of the angular momentum vector 
of the binary given by 
\begin{equation}
i = \cos^{-1}(\hat{\bm{l}}_{\rm b}\cdot \hat{\bm{l}}_{\rm p})\,,
\end{equation}
where $\hat{\bm{l}}_{\rm b}$ is a unit vector in the direction of the 
binary's angular momentum and $\hat{\bm{l}}_{\rm p}$ a unit vector in 
the direction the test particles angular momentum would be 
if it were not zero mass.
The nodal phase angle measured relative to the eccentricity 
vector of the binary is given by
\begin{equation}
    \phi = \tan^{-1}\left(\frac{\hat{\bm{l}}_{\rm p}\cdot (\hat{\bm{l}}_{\rm b}\times 
    \hat{\bm{e}}_{\rm b})}{\hat{\bm{l}}_{\rm p}\cdot \hat{\bm{e}}_{\rm b}}\right) + 90^\circ ,
\end{equation}
\citep{Chen2019,Chen2020e} where $\hat{\bm{e}}_{\rm b}$ is the eccentricity vector of the binary. 


\subsection{Comparison to previous work without 
General Relativistic Effects}
\label{nogr}

\newcommand{\icss}{$(i\cos\phi,i\sin\phi)$ surface}

We first ran some simulations without GR with a binary consisting of two stars with masses $m_1=m_2=0.5\,\rm M_\odot$ with  a semi-major axis of $a_{\rm b} = 0.5\,$au and eccentricities of 
$e_{\rm b} = 0.2$, 0.4, 0.6 and $0.8$.  The test particles began
in a circular orbit with a semi-major axis of $r=5 \,$au.  The orbits have an initial inclination, $i_0$, between  10 and 170$^\circ$ with an interval of $10^\circ$ and initial longitude of ascending node $\phi=90^\circ$.

Fig.~\ref{fig:nogr} shows the $(i\cos\phi,i\sin\phi)$ phase plane for these orbits.
In this diagram, an orbit 
with the angular momentum vector aligned, or anti-aligned, with 
the eccentricity vector of the binary is a point at 
(0,90), or at (0,-90) respectively.  A coplanar orbit is a point at (0,0) and a retrograde coplanar orbit is a circle of 
radius $180^\circ$ centered around (0,0).  In the case of the retrograde coplanar orbit,
the phase angle is not defined and so these orbits are represented
on the figures in black for reference purposes as they represent the edge of the phase space.  
The stationary inclination is the inclination about which the (red) librating orbits precess. For these orbits without GR it is always $i_{\rm s}=90^\circ$. 


As the eccentricity of
the binary increases we see the 
librating regions expand and the circulating regions
shrink.   The librating 
orbits are symmetric  about $i=90^\circ$ and so librating 
orbits with the initial inclinations
of the same offset from $i_{\rm s}=90^\circ$ in this plot trace over 
each other (e.g. $i_0=100^\circ$ traces over $i_0=80^\circ$ and so on).
The minimum initial inclination for librating orbits, $i_{\rm min}$, decreases with the binary eccentricity.
These plots are in good agreement
with the test particle simulations
in \cite{Doolin2011} as well as the low planet 
mass simulations of \cite{Chen2019}. 

We also ran these cases without GR for particle semi-major axes of $10$,
$20$, $30$ and $100\,$au  and found that these are nearly identical to the $r=5\,$au orbits. Note that 
while the 
track on the \icss\ is independent of the particle semi-major axis, the 
timescale for one complete precession
around the stationary inclination is not. Also, at small 
radii the orbits can become unstable due to interactions with
the binary \citep[e.g.][]{Hong2019,Chen2020}.



\subsection{With General Relativistic Effects at large radii}
\label{sec:withgr}

Next, we consider the effects of  GR on circumbinary particle orbits.
The simulations are run with the same initial conditions as described in the previous section but
we fix $e_{\rm b}=0.6$ and consider varying particle semi-major axes.
The results are shown in Figs.~\ref{fig:withgr1} \RGM{and~\ref{fig:withgr2}}.
Comparing the upper left panel of Fig.~\ref{fig:withgr1} to the lower left panel of Fig.~\ref{fig:nogr},  we see that the orbits at $r=5\,\rm au$ are nearly identical 
to those without GR.  

\begin{figure*}
\includegraphics[width=2\columnwidth]{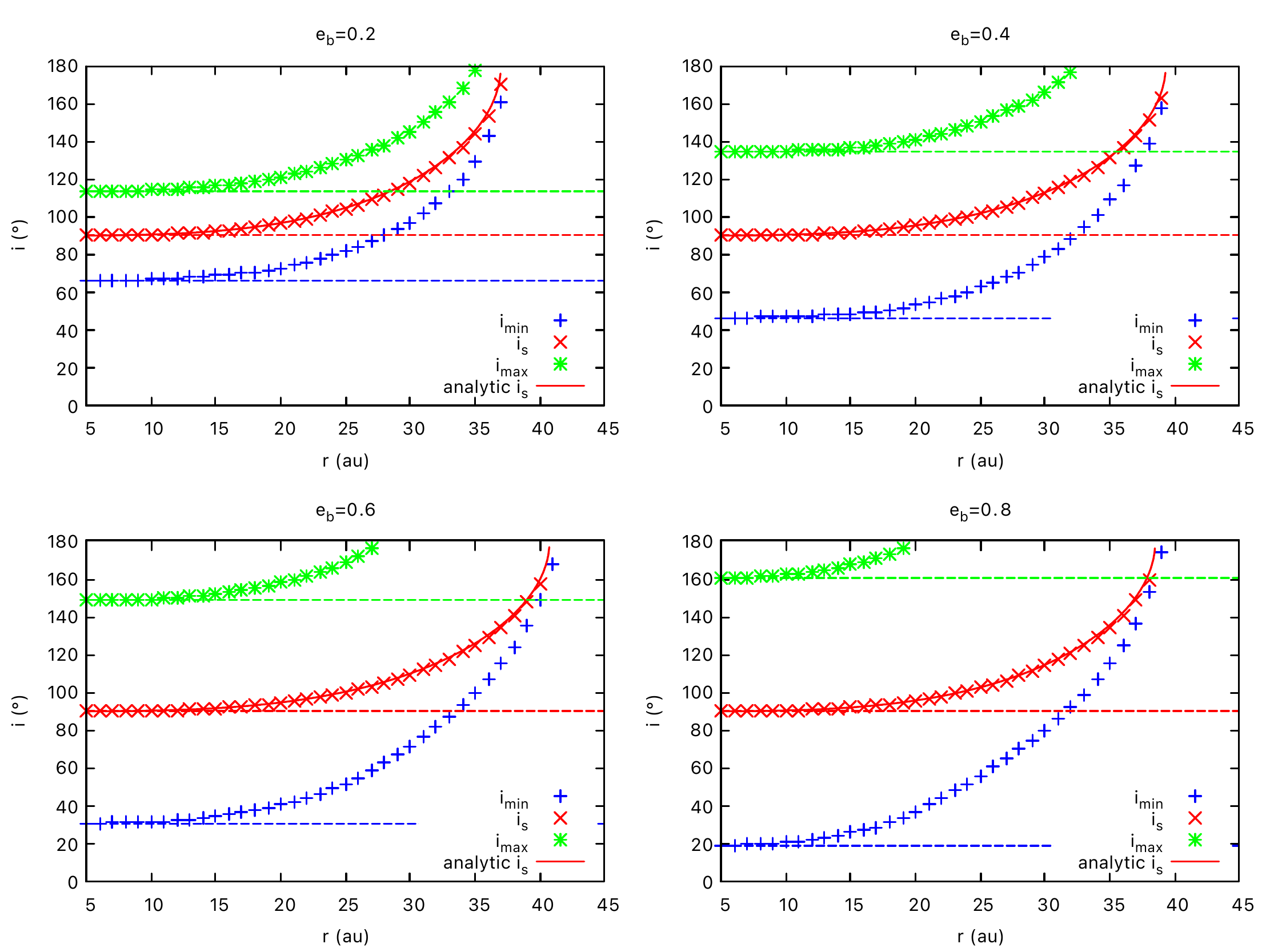}
\caption{The minimum initial librating inclination (blue), 
the stationary inclination inclination (red) and the maximum initial
librating inclination (green) vs semi-major axis  of the test particle orbit 
for an inner binary of $a_{\rm b}=0.5\,\rm au$ and $e_{\rm b}$ of 0.2 (top left), 0.4 (top right), 0.6 (bottom left) and 
0.8 (bottom right) including GR effects.  The solid red line shows the analytic expression for 
the stationary inclination (equation~\ref{eq:polar}).  Without GR 
the minimum librating inclination, the stationary inclination and 
the maximum librating inclination are independent of distance and 
shown as horizontal dashed lines in blue, red and green respectively. 
}
\label{fig:libreg}
\end{figure*}

As the particle semi-major axis increases,  the stationary inclination at the center of the librating region
moves to higher (retrograde) inclinations. 
The circulating region expands and the retrograde circulating region shrinks. 
One effect of the moving stationary inclination is 
that the librating orbits appear to have thicker lines in
the $r=10\,\rm au$ case compared with the $r=5\,\rm au$ case.   This is because the stationary inclination has moved slightly away from $90^\circ$. The 
initial inclinations relative to the stationary inclination are no longer symmetric about the stationary inclination.  
Therefore orbits beginning at say $80^\circ$ and $110^\circ$ no longer lie on top of each other.
Also note that the $150^\circ$ orbit has become librating in
the $r=10\,\rm au$ panel as the stationary inclination moved to
higher inclinations.
As we move to even larger 
radii, 
eventually the librating region 
disappears completely as the stationary inclination moves beyond
$180^\circ$.   
For particle semi-major axis greater than about $40\,\rm au$, there 
are only circulating orbits, as can be seen in the $r=42\,\rm au$ panel.
The effects of GR have removed the possibility for librating orbits
at large orbital radius.

\SL{Fig.~\ref{fig:withgr2} shows another view of the same results shown in Fig~\ref{fig:withgr1} except here the inclination and phase angle are plotted on the surface of a sphere.  The eccentricity unit vector and angular momentum unit vector are also plotted.  The angular momentum unit vector points at  the prograde circulating region, which in this view is on the back left of sphere and the eccentricity unit vector points, in the r=5 au  plot, at the librating region on top of the sphere.  At greater radius, $r$, both the librating regions move toward $180^\circ$ and merge.}

Next we explore the librating region further by plotting  in  Fig. \ref{fig:libreg} the stationary inclination, 
the lowest librating 
initial
inclination, $i_{\text{min}}$, and the highest 
librating initial inclination, $i_{\text{max}}$, as a 
function of particle radius for the binary eccentricities of $e_{\rm b}=0.2$,
$0.4$, $0.6$ and $0.8$.  The librating region is found 
by changing the inclination by one degree and finding the 
first inclination that librates and the last inclination that librates. 
The stationary inclination is found by searching which initial inclination
is stationary on the \icss.  This is done by starting with
inclinations on each side of the stationary inclination and integrating for 
two test particle orbital periods.  The projection of 
the orbits on the \icss\ move in opposite directions. On 
the y-axis an initial inclination less than the stationary inclination 
will initially move to smaller angles for $\phi$, while an initial
inclination larger than the stationary inclination will initially move to larger
angles.  If we bracket the stationary inclination a 
binary search root finder can be used to find the stationary inclination.

\begin{figure*}
\includegraphics[width=2\columnwidth]{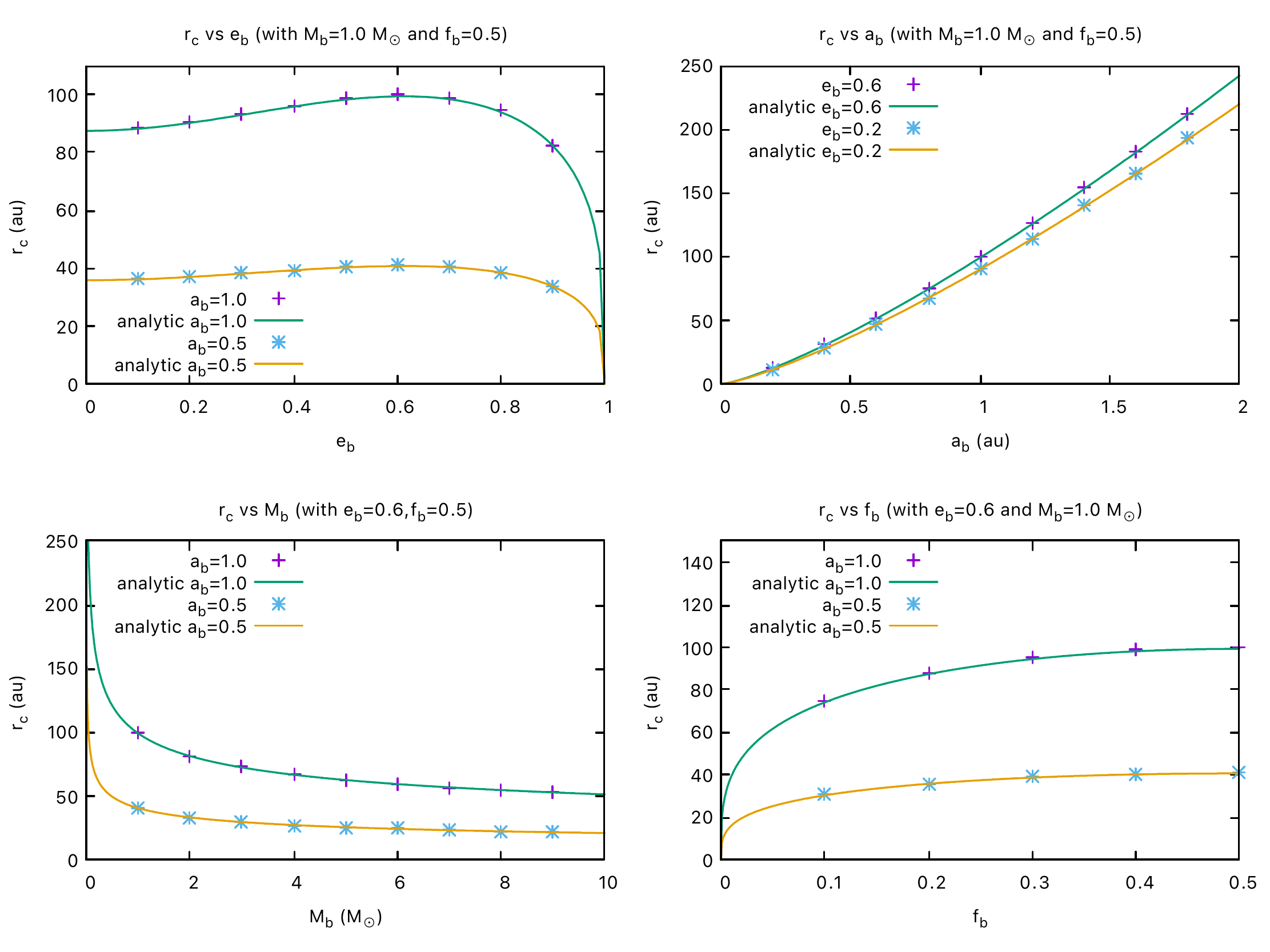}
\caption{The largest particle semi-major axis which has librating orbits, $r_{\rm c}$, as a function 
of binary eccentricity (top left), binary semi-major axis (top right), total binary mass (bottom left) and binary mass fraction (bottom right). 
}
\label{fig:critrad}
\end{figure*}

As with the \icss\ panels, the panels in Fig. \ref{fig:libreg}
map out the librating region.
At $r=5\,$au the librating region extends over the same range
as was found in the without GR calculations. The $e_{\rm b}=0.2$
case has the smallest librating region and the $e_{\rm b}=0.8$ case has the 
largest.  As the particle semi-major axis increases, the stationary inclination and the librating region tend 
toward $180^\circ$.  First to reach $180^\circ$ is the 
maximum inclination of the librating region.  At this point
there is no longer a retrograde circulating region.  Another 
interesting point is where the lower edge of the librating 
region passes $90^\circ$, at this point the librating region 
lies entirely in the retrograde region. For the $e_{\rm b}=0.2$ case 
this happens before the higher edge reaches $180^\circ$ and 
so for some radii the librating region is bracketed by prograde and retrograde 
circulation.  Eventually the stationary inclination reaches $180^\circ$ and no
more librating orbits are possible.

The orbit of a test particle around an inner binary including 
GR effects was explored by \cite{Naoz2017} and \cite{Zanardi2018}.  They used a quadrupole approximation and explored the effect
general relativity has on a test particle orbit. General relativity
causes the inner binary to precess on a timescale of 
\begin{equation}\label{eq:grprec}
    t_{GR} = 
    2\pi \frac{a_{\rm b}^{5/2}c^2 (1-e_{\rm b}^2)}{3k^3(m_1+m_2)^{3/2}},
\end{equation}
where  $c$ is the speed of light and $k^2$
is the gravitational constant.
This means that the frame of the binary is a rotating frame. 
They solve for the orbits within this rotating frame and find
how the ascending node of the outer planet changes with time.
\cite{Zanardi2018} take this result and find the
stationary inclination, $i_{\rm s}$, where the ascending node relative to the binary
does not change to be
\begin{equation}\label{eq:polar}
    i_{\rm s} = \cos^{-1}\left( A \frac{r^{7/2}}{a_{\rm b}^{9/2}
    (1-e_{\rm b}^2)(2+8e_{\rm b}^2)}\right) 
\end{equation}
(this is their equation (7) with the 
eccentricity of the test particle set to zero and the ascending 
node of the test particle to $-90^\circ$), 
where $A$ is a constant given by 
\begin{equation} 
A = -\frac{8 k^2(m_1+m_2)^3}{c^2 m_1 m_2} \,.
\end{equation}
This is plotted in Fig \ref{fig:libreg} as the solid red lines and 
is seen to be in good agreement with  our numerical results.

Another way to look at the stationary inclination, is 
to consider the calculation without GR as was shown in 
Fig. \ref{fig:nogr}.  Here we can see that the orbits in 
the \icss\ are such that when they cross the y-axis there 
is always an extemum of inclination.  So the inclination
is initially constant as the orbit moves off the y-axis and 
all the change comes from a change in $\phi$.  This change in
$\phi$ is toward smaller $\phi$ for inclinations less than $90^\circ$
and toward larger $\phi$ for inclinations greater than $90^\circ$.
The inclination doesn't change at a uniform rate, but if one 
takes the average over a full test particle orbit then one
can get an average rate of change for a particular radius and 
initial inclination.
The precession of the binary due to GR is such that it causes a 
change in $\phi$ toward smaller values.  Thus, there is some 
inclination greater than $90^\circ$ where on the y-axis in 
the \icss\ the precession from GR exactly cancels the motion 
in $\phi$ from the initial motion of the orbit caused by the 
interaction of the binary and the test particle orbit.  We 
have written an addition program, which for each radius of 
test particle orbit in a non-GR calculation, 
it searches for an inclination in which the initial motion 
of $\phi$ averaged over two test particle orbits is equal to 
the precession which would be caused by GR and this reproduces
our stationary inclination calculated with GR.

\section{Critical radius for librating orbits}
\label{critical}

Without GR, the stationary inclination 
is at $90^\circ$ to the binary orbit and is independent of the particle semi-major axis (see Section~\ref{nogr}).  With GR, there is a critical semi-major
axis, $r_{\rm c}$,  where the stationary inclination reaches $180^\circ$ and beyond this there
are no librating orbits (see section~\ref{sec:withgr}).
We found this radius numerically in Fig.~\ref{fig:critrad} for varying binary eccentricity, semi-major axis, total binary mass, $M_{\rm b}$ and binary mass fraction $f_{\rm b}=m_2/M_b$.


We can 
also solve for the critical radius, $r_{\rm c}$, by setting
the stationary inclination to $180^\circ$ in Equation \ref{eq:polar}
and solving for the critical semi-major axis to get 
\begin{equation}
    r_{\rm c}=a_{\rm b}   
    \left( \frac{-a_{\rm b}}{A} (1-e_{\rm b}^2)(2+8e_{\rm b}^2) \right)^{2/7}
    \,.
\end{equation}
For some typical parameters we can write this as
\begin{equation}
    r_{\rm c} = 40.7 {\,\rm au} 
    \left(\frac{a_{\rm b}}{0.5\,\rm au}\right)^{9/7} 
    \left(\frac{M_{\rm b}}{1\,M_\odot}\right)^{-2/7} F_1(f_{\rm b}) F_2(e_{\rm b}),
\end{equation}
where $F_1(f_{\rm b})$ and  $F_2(e_{\rm b})$ are functions 
with a maximum value of one characterizing the effects of 
the mass fraction and eccentricity on the critical semi-major axis.
They are given by
\begin{equation}
    F_1(f_{\rm b})=\left(4f_{\rm b}(1-f_{\rm b})\right)^{2/7}
\end{equation}
    and
    \begin{equation}
    F_2(e_{\rm b})=\left(\frac{8 (1-e_{\rm b}^2)(2+8e_{\rm b}^2)}{25}\right)^{2/7},
\end{equation}
where the maximum value for these functions occur at $f_{\rm b}
=0.5$ and $e_{\rm b}=\sqrt{3/8} \approx 0.6$, respectively.
These analytic results are plotted also in Fig \ref{fig:critrad}
along with the numerical simulations discussed above 
and they match very well.  

The critical radius goes to infinity, matching the 
non-GR case as $a_{\rm b}$ goes to infinity and as $M_{\rm b}$ 
goes to zero.  This can be explained as in both these limits, 
the binary precession rate goes to zero which matches the lack of
precession in the non-GR case.  As the $a_{\rm b}$ goes to 
zero or the $M_{\rm b}$ goes to infinity the precession rate 
goes to infinity and the critical semi-major axis goes to zero.
The critical semi-major axis is not a strong function of either 
binary eccentricity  or mass fraction.  However,  the critical
semi-major axis goes to zero as the binary
eccentricity goes to 1, where the precession rate also goes 
toward infinity,  In the case of the mass fraction, the 
critical radius goes to zero as the mass fraction goes to zero, but
this is because the binary is no longer a binary in this limit.

Finally, we examine the radius at which the stationary inclination
reaches a particular angle.  The stationary inclination varies from 
the non-GR case of $90^\circ$ to the limit of $180^\circ$ which
it reaches at $r_{\rm c}$.  The radius or semi-major axis where it 
passes through a particular angle is a fixed fraction of the 
critical semi-major axis.  This fraction is given by 
\begin{equation}
    F_3(i)=(-\cos(i))^{2/7} \,.
\end{equation}
So for example the stationary inclination will be at $i_s=95^\circ$
at a radius of $F_3\approx 0.50$ of the critical radius and the stationary inclination will be at $i_s=110^\circ$ 
when the radius is at $F_3\approx 0.74$ of $r_{\rm c}$.

\section{Conclusions}
\label{conc}

Circumbinary orbits around an eccentric binary undergo nodal precession. The precession may be centered on the binary angular momentum vector (circulating orbits) or the stationary inclination (librating orbits). In the absence of general relativity, the stationary inclination is always at $90^\circ$ for a test particle and aligned with the binary eccentricity vector. We have shown that the effects of general relativity increase the stationary inclination. For particles that are close to the binary, the stationary inclination remains close to $90^\circ$, however it increases with the particle semi-major axis.
There is a critical radius outside of which there are no polar librating orbits, only circulating orbits.   The critical radius is not very sensitive to the binary eccentricity but increases with the binary semi-major axis. The critical radius is in planet forming regions around close binaries with separation $a_{\rm b}\lesssim 1\,\rm au$.

We have not included the effect of the mass of the particle in our calculations. A massive third body causes a retrograde apsidal  precession of the binary. If the particle is massive enough, it may effectively cancel out the prograde precession caused by general relativity. However, the work presented here does not change significantly if the outer test particle is replaced by a Jupiter mass planet \citep[see also][]{Chen2019,Martin2019} and does not affect the orbit of the binary \citep{MartinD2016}. However, the effect of the mass may become important when considering a massive or radially extended circumbinary disk \citep[see also][]{Migaszewski2011}.

\SL{The particle simulations are run with point masses and are valid for that limit.  At small separations, \RGM{but before the orbits are circularised}, tidal forces and the spin of the stars cause precessions which may compete with the GR precession. The precession due to these other processes can be quite complicated and depends on the exact composition and structure of the two stars \citep{sterne39,sirotkin09} or compact objects \citep{hamilton21}. } \RGM{A faster prograde precession would lead to a smaller critical radius for polar orbits to exist.}

\RGM{The dynamics of a circumbinary gas disk are qualitatively similar to those of test particles. Each ring of the disk wants to behave in the same way as a particle, however, communication between the rings leads to differing behaviour. For protoplanetary discs, the communication typically occurs in the bending waves regime \citep{PP1983}. The disk  remains  in a coherent structure if the communication timescale (that is about half the sound speed) is shorter than the precession timescale \citep{Papaloizou1995,Larwoodetal1996,Papaloizou1998}. In this case, a given radius of the disc may display dynamically different behaviour to a particle at the same radius. However, circumbinary disks can be sufficiently radially extended that they are not in good communication and can undergo disk breaking  \citep{Nixon2013}.} 

\RGM{Another difference with the disk compared to the particles is that there is viscous dissipation.} A  misaligned circumbinary gas disc moves towards either coplanar or polar alignment \citep{Papaloizou1995,Nixon2013,Martin2017,Martin2018}. \RGM{For a binary with a critical radius, $r_{\rm c}$, that is within a circumbinary disk, the evolution may depend on the radial extent of the disk and disk properties such as the sound speed and the viscosity.   }
If the disc extends beyond $r_{\rm c}$, disc breaking may lead to an inner polar disc and an outer disc that is circulating and moving towards coplanar. It is interesting to note that the  observed polar gas disc HD~98800 has a exterior companion that could truncate the circumbinary disc and prevent it from spreading to the critical radius that we have found \citep{Martin2022}. The semi-major axis of HD98800 is about $1\,\rm au$ \citep{Kennedy2019} 
and this corresponds to a critical radius of about $r_{\rm c}=80\,\rm au$, 
much larger than the size of the observed disc.  

These results have important implications for observations of misaligned circumbinary planets.
Stationary polar planets and polar librating planets, can only be inside of the critical radius. Similarly, debris discs, that may typically extend to tens or hundreds of au, 
must be close to the binary to be polar-aligned.
Note that the polar debris disc around 99 Herculis would not be affected by GR since the separation of the binary is about $16.5\,\rm au$ \citep{Kennedy2012,Smallwood2020}, much larger than those considered in this work.  This suggests that we would not expect a solar system analogue with a polar close central binary. The inner planets in the solar system could be polar, but the outer planets and the Kuiper belt or debris disc are more likely to be coplanar if they are the remnants of a circumbinary gas disc.

\begin{acknowledgements}
 We acknowledge support from NASA through grant 80NSSC21K0395.  
 We would like to thank the referee for helpful comments.
\end{acknowledgements}

%
%
%
%

\end{document}